\documentstyle[12pt]{article}
\begin{document}

\topmargin 0pt

\oddsidemargin -3.5mm

\headheight 0pt

\topskip 0mm
\addtolength{\baselineskip}{0.20\baselineskip}
\begin{flushright}
MIT-CTP-$2750$
\end{flushright}
\begin{flushright}
hep-th/9806119
\end{flushright}
\vspace{0.5cm}
\begin{center}
    {\large \bf  Statistical Entropy of  Three-dimensional Kerr-De Sitter 
Space}  
\end{center}
\vspace{0.5cm}
\begin{center}
 Mu-In Park\footnote{Electronic address: mipark@physics.sogang.ac.kr 
{\it and} mipark@ctpa03.mit.edu} \\
{ Center for Theoretical Physics and Laboratory for Nuclear Science, \\
Massachusetts Institute of Technology, Cambridge, Massachusetts 02139 U.S.A.}
\end{center}
\vspace{0.5cm}
\begin{center}
    {\bf ABSTRACT}
\end{center}
I consider the (2+1)-dimensional Kerr-De Sitter 
space and it's statistical entropy computation.
It is shown that this space has only one (cosmological) event horizon and 
there is a phase transition between the stable horizon and the evaporating 
horizon
at a point $M^2=\frac{1}{3}J^2/l^2$ together with a lower bound of the horizon 
temperature.
Then, I compute the statistical entropy of the space by using a recently 
developed formulation of Chern-Simons theory with boundaries, and extended 
Cardy's formula. This is in agreement with the thermodynamics formula.
\vspace{3cm}
\vspace{2.5cm}
\begin{flushleft}

15 June 1998 \\
\end{flushleft}

\newpage

\begin{center}
{\bf I. Introduction}
\end{center}

Recently, there has been tremendous interests in the statistical origin of the 
entropy for the (2+1)-dimensional space with a negative cosmological constant 
which is asymptotically the Anti-de Sitter space ($AdS_3$) [1]. Independently 
on the string paradigm, two typical approaches have been known: One is 
Carlip's approach [2] and the other is Strominger's one [3]. In the Calip's 
approach, one uses several assumptions. A first is the existence of a well 
defined conformal field theory with the Kac-Moody current algebra and it's 
related Virasoro algebra through the Sugawara construction. A second is the 
appropriate boundary conditions which yield the desired value for the entropy.
Third, one assumes that all statistical degrees of freedom of black hole live
{\it on~the~black-hole~
event~horizon}.

The Strominger's approach is a 
drastically different one which concerns the Brown-Henneaux's  asymptotic 
isometry group $SO(2,2)$ [4] which preserves the 
asymptotic metric $g^{\mu \nu}$ of $AdS_3$. In this 
approach, the fact that there is a central charge as $c=12 l$ even at the 
``classical'' level is a basic  ingredient. [Here, Newton's constant is set 
$G \equiv 1/8$ and cosmological constant is $\Lambda =-1/l^{2}$.] 

However, it was not 
clear how these two extremal approaches are connected.
Involved with this problem, recently Ba\~nados, Brotz and Ortiz (BBO) have 
considered the 
Chern-Simons gravity theory with boundaries (finite or infinite) [5, 6]. 
(See also Ref. [7] for recent compact review and comparison with other 
various formulations.) In the 
Chern-Simons theory with boundaries, there are Kac-Moody and it's related 
Virasoro algebras with the central terms at the ``classical'' level, which 
was 
first argued \footnote{ 
For the Kac-Moody algebra, it was known in a different context of Yang-Mills 
theory with Chern-Simons term in Ref. [8].}
by Ba\~nados [5] and recently proved [9] in the symplectic 
method [10] and this algebra has a crucial role in their formulation. Their 
formulation produces, `` independently on the radius of the outer boundary, 
which envelopes all the space'', 
the Bekenstein-Hawking's thermodynamics entropy for the BTZ black hole [1]
\begin{eqnarray}
S=2 \pi \sqrt{l (l M +J)} +2 \pi \sqrt{l (l M -J)}  
\end{eqnarray}
with black hole mass $M$ and angular momentum $J$ [1, 2, 3]. In this 
derivation, it is a basic ingredient that the central charge of the Virasoro 
algebra is completely determined by matching the isometries asymptotically; 
The central charge is found to be the same as that of asymptotic isometries.

Now, with this powerful formulation, the previous two extremal approaches can 
be understood as some limiting cases. Moreover, it provides a simple answer 
about the reason for the same result of the two previous approaches: They 
treated 
an identical object which lives only on the boundary ! However, contrast to 
the $\Lambda < 0$ case, the analysis of statistical entropy for the 
(2+1)-dimensional space with $\Lambda >0$ which is asymptotically De Sitter 
space ($DS_3$) [11, 12], has not been well studied. Actually, the $\Lambda >0$
case is quite different to the $\Lambda <0$ case. In the $DS_3$ space which 
is the simplest case of $\Lambda >0$
\begin{eqnarray}
ds^2_{}=-\left(1 -\frac{r^2}{l^2} \right) dt^2 
+\left(1-\frac{r^2}{l^2} \right)^{-1}dr^2 +r^2 d \varphi^2 
\end{eqnarray}
there is no black-hole event horizon for an observer moving on a timelike 
world line, but there is a cosmological event horizon $r_+ =l$ separating the 
outside region which the observer can never see from the inside region that he
can see if he waits long enough. This is an opposite situation to the usual 
black hole spacetimes. However, as have been shown by Gibbons and Hawking, the
cosmological event horizon has many formal similarities with the black-hole 
event horizon. Furthermore the ideas of thermodynamics for the black-hole 
event horizons whose areas can be intepretated as the entropies to the 
cosmological event horizons, but by abandoning the concept of particle as 
being observer-independent 
[11].
For the special case $DS_3$, the statistical analysis was recently done
by 
Maldacena and Strominger [13] by applying the Carlip's approach instead of the
Strominger's one. In this analysis they were able to show a good agreement 
with the Gibbons-Hawkings formula.
However, there remains some gap to the complete understanding of the 
statistical entropy for {\it far-horizon} region  and more general cases with 
$M$ and $J$.
 
In this paper, I consider a Kerr-De Sitter space with the general $M, J$ with 
$\Lambda > 0$ and a computation of it's statistical entropy. Following the 
Gibbons-Hawking's approach, it is found that this space has only one 
(cosmological) event horizon and there is a phase transition between a stable 
horizon and an (unstable) evaporating horizon at a point 
$M^2=\frac{1}{3}J^2/l^2$ together
with a lower bound of horizon temperature. Then,
I compute the statistical entropy of the space by a direct adaptation of the 
approach of BBO [5 - 7, 9] and extending the Cardy's formula to the complex 
valued central charge and eigenvalues of $L_0,~\bar{L}_0$.
My result agrees with the thermodynamics formula exactly.

\begin{center}
{\bf II. Kerr-$DS_3$ solution}
\end{center}

In order to proceed parallel to $\Lambda <0$ case [1], I start by considering 
the Kerr metric in $\Lambda >0$ case [11, 12]. The (2+1)-dimensional gravity 
for a cosmological 
constant $\Lambda =-1/l^{2}$ is described by the action
\begin{eqnarray}
I=\frac{1}{2 \pi} \int d^3 x \sqrt{-g} ( R +2 l^{-2})+I_m, 
\end{eqnarray}
where $I_m$ is a presummed matter action [The details are not important in 
this paper] and I have omitted the surface terms in the pure gravity part as 
usual. This theory has a constant 
curvature $R=-2 l^{-2}$ outside the matters. Regardless of 
the sign of $l^{2}$, the vacuum line element for the rotationally symmetric 
and 
stationary metric can be written as
\begin{eqnarray}
ds^2 =-(N_{\bot})^2(r) dt^2 +f^{-2}(r) dr^2 +r^2 (N^{\varphi}(r) dt +d 
{\varphi})^2,
\end{eqnarray}
where $\varphi$ has period $2 \pi$. Then, the Hamiltonian is expressed by 
$H=\int dr (N {\cal H} +N^{\varphi} {\cal H}_{\varphi} )$ with the 
constraints
\begin{eqnarray}
{\cal H} &\equiv &- 2 l^2 ~\frac{p^2}{r^3} +(f^2)' +\frac{2r}{l^2} \approx 0, 
 \nonumber \\
{\cal H}_{\varphi} &\equiv &-2 i l p ' \approx 0, \nonumber \\
N(r) &= &f^{-1} N_{\bot},
\end{eqnarray}
where $N, N^{\varphi}$ are the Lagrange multipliers and prime $(~'~)$ denotes 
the derivative with respect to the radial coordinate $r$. The 
solutions of (5) depend on the sign of $l^2$. In my interesting case of 
$l^2 <0$, the solutions of $p, f^2, N_{\bot}, N^{\varphi}$ are given as follow:
\begin{eqnarray}
p&=&-\frac{J}{2 il},\nonumber \\
f^2 &=&N^2_{\bot}=M-\left(\frac{r}{l}\right)^2 +\frac{J^2}{4 r^2}, \nonumber \\
N^{\varphi} &=&-\frac{J}{2 r^2},
\end{eqnarray}  
where I have renamed $l$ by $il$ such that $l$ is a positive real number and 
I have set $N|_{\partial D_2}=1,~N^{\varphi}|_{\partial D_2}=0$ in order to
get the $DS_3$ 
space (2) asymptotically [1]. Here, two constants of integration $J$ and $M$, 
which characterize a $Kerr-DS_3$ space,
are identified as the total angular momentum and mass because they appear as 
the conjugates to the boundary (rescaled) lapse and shift displacements 
$N|_{\partial D_2}$ and $N^{\varphi}|_{\partial D_2}$, respectively, in the 
variation of the action (3) with an appropriate boundary action [1, 14]. 
\footnote{This definition of $M$ and $J$ is valid even for finite space. 
Moreover, $M$ and $J$ converge into the quasi-local or ADM definitions of mass
and angular momentum asymptotically [14] though the gravitational energy 
vanishes. There are other several methods of identifying the mass and angular 
momentum. See Ref. [15] for these other methods. I thank Prof. S. Carlip for 
suggesting me to consider this problem.}
Note that there is an additional sign change in front of 
$M$ as well as the $l^2$ terms. In this paper I will focus mainly on the 
statistical entropy of the solution (4), (6). The geometric structure is not 
main issue for this purpose and will not be provided in this paper.

The lapse function $N_{\bot}$ vanishes for ``one'' value of $r$ given by
\begin{eqnarray}
r_{+} =\frac{l}{\sqrt{2}} \sqrt{M+\sqrt{M^2 +\frac{J^2}{l^2}}}.
\end{eqnarray}
This is the cosmological event horizon in $Kerr-DS_3$ and there is no 
black-hole event horizon. Here, there is no additional condition for $M$ in 
order that the horizon exists unless $J$ vanishes: Even the negative values of
$M$ and $J$ are allowed. So, in the $J \neq 0$ case the whole mass spectrums 
(ranging form $-\infty$ to $\infty$) are continuous and there is no mass gap; 
This is contrast to $Kerr-AdS_3$ called $BTZ$ solution [1]. For $J=0$ case, 
there is no horizon when $M < 0$; One is left just with the outside region 
which is filled with negative masses.
Moreover, I define $r_- \equiv \frac{l}{\sqrt{2}}\sqrt{M-\sqrt{M^2 
+\frac{J^2}{l^2}}}
\equiv i r_{(-)}$ which is a pure imaginary number. With these two 
parameters, the $Kerr$ metric (4), for a positive cosmological constant 
$1/l^2$, can be conveniently written in the proper radial 
coordinates as
\begin{eqnarray}
ds^2 =\mbox{sinh}^2 \rho \left( \frac{r_+ dt}{l} -r_{(-)} d {\varphi} \right) 
^2 -l^2 d \rho^2 +\mbox{cosh}^2 \rho \left( \frac{r_{(-)} dt}{l} +r_+ 
d \varphi \right )^2
\end{eqnarray} 
with
\begin{eqnarray}
M=\frac{r_+^2 -r_{(-)}^2}{l^2}, ~ J=\frac{2 r_+ r_{(-)}}{l}, \nonumber \\ 
\nonumber \\
r^2=r^2_+ \mbox{cosh}^2 \rho + r^2_{(-)} \mbox{sinh}^2 \rho.
\end{eqnarray}
In these coordinates, the cosmological event horizon is at $\rho =0$ and hence
this metric 
represents the exterior of the horizon for real value $\rho$ and represents
the interior for imaginary value $\rho$. [This is completely opposite 
situation
to Schwarzshild black hole.] The interior and exterior regions are casually 
disconnected and so the cosmological event horizon acts like as a black-hole 
horizon 
\footnote{Because of this fact, I posit that the sources of $M$ and $J$ are 
isotropically distributed matters within cosmological horizon and outer 
boundary in accordance with black hole analogy where the sources hide also 
inside the (black-hole) event horizon. The centrifugal terms will be a result 
of Mach effect for the observer surrounded by the rotating mass shell [16].
Of course, my calculation of statistical entropy is independent on the 
precise physical setting for the metric solution (4), (6). However, if one 
accepts this interpretation, the BTZ black hole can be also intepretated as 
the rotation of the space filled with isotropically distributed $negative$ 
mass matters.}.
Here, I note that the sign of $M$ is controlled by the relative magnitudes of 
$r^2_{+}$ and $r^2_{(-)}$.
By considering $J=0$ case, the metric (4) can be identified with the $DS_3$ 
space (2) [11, 12] with $M=1$. 

Finally, let me applies the Gibbons-Hawking's thermodynamics theory to the 
interior region of $Kerr-DS_3$ where the signature of metric is the same as 
ours. Then, the observers in the interior will calculate the 
Bekenstein-Hawking entropy [11]
\begin{eqnarray}
S&=& 2 \cdot {\mbox Area~of~event~horizon} \nonumber \\
&=&4 \pi r_+
\end{eqnarray}
and detect an isotropic \footnote{ This implies that the concept of particle 
is observer dependent [11].} background of thermal radiation with a temperature
\begin{eqnarray}
T=\left(\frac{ \partial S}{ \partial M }\right)^{-1}_J 
=\frac{r_{+}^2 +r_{(-)}^2}{ 2 \pi l^2 r_+ }.
\end{eqnarray}
Moreover, according to an semiclassical argument of Gibbons-Hawking, a 
stability of the cosmological event horizon can be analyzed by considering 
the change
of temperature $T$ upon varying $M$ or in a more compact way by the heat 
capacity $C_J \equiv (\partial M/ \partial T)_J$ (with $J$ fixed) [14]: From 
(11), one obtains
\begin{eqnarray}
C_J=2 \sqrt{ 2} \pi l ~\sqrt{M} ~\frac{ \sqrt{1+x} ~(1+ \sqrt{1+x} )^{3/2} }
{1-x +\sqrt{1+x}},
\end{eqnarray}
where $x=J^2/(M^2 l^2)$ is a dimensionless parameter. This shows an infinite 
discontinuity at the point $M^2 =\frac{1}{3}J^2/l^2~(\mbox{i.e.,}~ x=3)$ and 
shows a 
critical phenomena; A physical 
interpretation is as follow: For $J=0, M >0$ case, if one absorb the thermal 
radiation at the expense of the mass of the horizon, the area of the horizon 
$( 2 \pi r_+)$ will go down, $T (={r_+}/({2 \pi l^2}))$ goes down 
($C_J =4 \pi l \sqrt{M} >0$), and hence the (cosmological) 
horizon is stable. On the other hand, for $J \neq 0, M>0$ case, as one 
absorbing the radiation from horizon, $T$ goes down ($C_J >0$) for 
$M^2 > \frac{1}{3} J^2/l^2~(\mbox{i.e.,}~x<3)$ but $T$ goes up ($C_J <0$) for
$M^2 < \frac{1}{3} J^2 /l^2~(\mbox{i.e.,}~x>3)$. From the fact that $C_J >0$ 
and/or 
$C_J <0$ 
imply the 
stability and/or instability of the horizon, one finds that there is a phase 
transition between the stable horizon and evaporating (unstable) event 
horizon at the
critical point. \footnote{ This looks like a second-order phase transition in 
the usual (equilibrium) thermodynamics because of an (infinite) discontinuity
in the second derivatives of the Gibbs free energy $G=M-TS -\Omega J~(\Omega 
\equiv - T (\partial S/ \partial J)_M)$ even though $G$ and it's first 
derivatives are continuous. Similar phenomena have observed also in the 
(3+1)-dimensional Kerr-Newmann black holes [20]. However, according to a 
recently developed {\it non-equilibrium} thermodynamics there is the 
second-order phase transition at $M=0, J=0$ point, but not at 
$M^2=\frac{1}{3} J^2/l^2$ and corresponding critical exponents satisfy the 
scaling laws [21]. The details will be appeared in a separate paper [22].} 
This is contrast to the BTZ black-hole event horizon which is
always stable [14] and the Scwarzschild black-hole horizon which always 
evaporate upon thermal radiation in the vacuum. Moreover, in this case there 
is also a lower bound of temperature as 
\begin{eqnarray}
&&T \geq T_c ~, \nonumber \\ 
&&T_c=\sqrt{\frac{2}{ 3 \pi l}}~ \sqrt{M},
\end{eqnarray} 
where $T_c$ is the horizon temperature at the critical point, which is lower 
than the temperature for the extremal point $M^2=J^2/l^2~(x=1)$.

\begin{center}
{\bf III. Chern-Simons gravity with boundaries}
\end{center}

The (2+1)-dimensional pure gravity with the positive cosmological constant 
$\Lambda =2 l^{-2}$ can be written as a $SL(2, {\bf C})$ Chern-Simons gauge 
theory [17, 18]. The action for this theory is, up the surface terms 
\footnote{Recently, Ba\~nados and Mendez proved that the surface terms in the 
covariant form of Chern-Simons gravity action like as (14) are exactly the 
same as the required surface terms in the pure gravity action [19].}, 
\begin{eqnarray}
I_g [A]=\frac{is}{4 \pi} \int_{D_2 \times R} d^3 x  \epsilon^{\mu \nu \rho} 
\left<A_{\mu} \partial _{\nu} A_{\rho} +\frac{2}{3} A_{\mu} A_{\nu} A_{\rho} 
\right>
-\frac{is}{4 \pi} \int _{D_2 \times R} d^3 x \epsilon^{\mu \nu \rho} 
\left< \bar{A}_{\mu} \partial _{\nu} \bar{A}_{\rho} +\frac{2}{3} \bar{A}_{\mu}
 \bar{A}_{\nu} \bar{A}_{\rho} \right>
\end{eqnarray}
on the manifold 
$\Sigma=D_2 \times R$. [$D_2$ is a 2-dimensional disc of space and 
$R$ is a 1-dimensional infinite real manifold of time. $\bar{A}_{\mu}$ is 
complex conjugate of $SL(2, {\bf C})$ gauge field 
$A_{\mu}$ and $ \left< \cdots \right>$ denotes the trace.] Here, the 
topological mass parameter `$s$' needs not be quantized in the non-compact 
group 
$SL(2, {\bf C})$ irrespective of the existence of the boundaries. 
Action (14) can be 
identified to (3) by the gauge connections \footnote{I 
take $
t_0=\frac{1}{2} \left(\begin{array}{cc} 0 & -1 \\ 1 & 0 \end{array} \right)
,~ t_1=\frac{1}{2} \left(\begin{array}{cc} 1 & 0 \\ 0 & -1 \end{array} 
\right),~ t_2 =\frac{1}{2} \left( \begin{array}{cc} 0 & 1 \\ 1 & 0 \end{array}
 \right)$ so that 
$[t_a, t_b]={\epsilon_{ab}}^c t_c$ and $\left< t_a t_b \right> =\frac{1}{2}
\eta_{ab}$, 
where $\epsilon_{012}=1$ and $\eta_{ab}=diag(-1,1,1)$. These are the same 
conventions as Ref. [6].}
\begin{eqnarray}
A^a_{\mu}=\omega^a_{\mu}+\frac{e^a_{\mu}}{i l},~
\bar{A}^a_{\mu}=\omega^a_{\mu}-\frac{e^a_{\mu}}{i l} ~~(a=0,1,2),
\end{eqnarray}
with $s=-l$.
Here, $e^a=e^a_{\mu}d x^{\mu},~ \omega^a=\frac{1}{2} \epsilon^{abc} 
\omega_{\mu b c} d x^{\mu}$ are the triads and the $SL(2, {\bf R})$ spin 
connections, respectively. From now on, I will only consider the 
$A_{\mu}$-part, unless otherwise stated, because the $\bar{A}_{\mu}$-part can 
be 
obtained by complex conjugation of $A_{\mu}$-part. It is easily checked that 
the 1-form gauge 
connections are given, in the proper coordinates, by
\begin{eqnarray}
{\bf A}^0& =&-\frac{r_+ +i r_{(-)}}{l} \left( \frac{dt}{l} +i d {\varphi} 
\right) \sinh \rho \nonumber \\
{\bf A}^1 &=& d \rho, \nonumber \\
{\bf A}^2 &=&-\frac{r_+ +ir _{(-)}}{l} \left( \frac{dt}{l} 
+i d \varphi \right) \cosh \rho.
\end{eqnarray}
[ The superscript indices denote the group indices $a=0,1,2$.]
For the $DS_3$ space (2), these reduce to
\begin{eqnarray}
{\bf A}^0 &=&\mp \sqrt{1 -\frac{r^2}{l^2}} \left(i \frac{dt}{l} -d \varphi 
 \right), \nonumber \\
{\bf A}^1 &=&\mp \frac{i}{\sqrt{l^2-r^2}} dr, \nonumber \\
{\bf A}^2 &=&\frac{r}{l^2} dt +\frac{ir}{l} d \varphi
\end{eqnarray}
using the coordinates $(t,r,\varphi)$ and on-shell mass $M=1$ [13]. In 
general cases, (16) becomes, in matrix form,
\begin{eqnarray}
{\bf A} =\frac{1}{2} \left( 
\begin{array}{cc}
 d \rho & -{z}~ e^{-\rho} \left( \frac{dt}{l} +i d \varphi \right) \\
 -{z}~ e^{\rho} \left( \frac{dt}{l} +i d \varphi \right) & - d \rho 
\end{array}\right),
\end{eqnarray}
where ${z}\equiv(r_+ +i r_{(-)})/l$. From this, the polar components 
\footnote{Here, $A_{\rho}=\hat{\rho}^i A_i, A_{\varphi} =\hat{\varphi}^i A_i$, 
for the orthogonal unit vectors $\hat{\rho}, \hat{\varphi}$ on 
$\partial D_2$.} 
in the 
proper coordinates can be obtained as 
\begin{eqnarray}
A_{\rho}=t_1,~A_{\varphi}=-i{z} ~({U}^{-1} t_2 {U} ),~ A_t =i A_{\varphi},
\end{eqnarray}
where \footnote{In this derivation, the additional gauge fixing conditions 
are not needed. See Refs. [5. 6] for comparison.}
\begin{eqnarray}
{U} = \left(
\begin{array}{cc}
e^{\rho /2} &0 \\
0 & e^{-\rho/2}
\end{array} \right).
\end{eqnarray}

\begin{center}
{\bf A. Symmetry algebra and  classical central terms}
\end{center}

The Chern-Simons action has the gauge and diffeomorphism ({\it Diff}) 
symmetries.
If there are boundaries, the central terms appear in the symmetry algebras 
even at the classical level. This was first argued by Ba\~nados 
[5] and proved recently [9] in the symplectic method [10]. 
Especially, for the time-independent and spatial {\it Diff}
\begin{eqnarray}
\delta_f x^{\mu} &=&-\delta^{\mu}_{~ k} f^k, \nonumber \\
\delta_f A^a _i &=&f^k \partial_k A^a_i +
(\partial_i f^k) A_k^a, \nonumber \\
\delta_f A^a _0 &=&f^k \partial_k A^a_0,
\end{eqnarray}
the conserved Noether charge becomes
\begin{eqnarray}
Q(f) = -\frac{i s}{4 \pi}
\oint _{\partial D_2} d \varphi ~\eta_{ab} 
(2 f^{\rho} A^a_{\rho} A^b_{\varphi} +f^{\varphi} A^a_{\varphi}A^b_{\varphi}
+f^{\varphi} A^a_{\rho} A^b_{\rho} ),
\label{cent}
\end{eqnarray}
where a boundary condition 
``$ A^a_{\rho} |_{\partial D_2} =\bar{A}^a_{\rho} |_{\partial D_2}$=
constant '' is imposed, $A_{\varphi}$ is a pure gauge form 
$A_{\varphi} =g^{-1}\partial_{\varphi} g$, $f^k$ is 
a real function of spatial coordinates and the boundary ${\partial D_2}$ is a 
circle. [The last constant term in (22) was included to obtain the standard 
Virasoro central term, which procedure can be always 
done according to the definition of Noether charge.] Here, the boundary 
condition about $A_\rho$ is crucial for the existence of central term in the 
Virasoro algebra but, in a general text, one can not discard other boundary 
conditions which do not produce the central term [9]. However, in our analysis
of spacetime with event horizons, it is quite natural choice according to the 
solution (19) [5 - 7]. From the symplectic 
structure of the action (14) for the pure gauge $A_i=g^{-1}\partial_i g$, one 
finds the Poisson bracket algebra for $A_{\varphi}^a$ who lives on 
$\partial D_2$:
\begin{eqnarray}
\{A_{\varphi}^a(\varphi), A_{\varphi}^b (\varphi ') \}
& =&\frac{2 \pi} {is}
 {\epsilon^{ab}}_{c} A_{\varphi}^c 
(\varphi) \delta(\varphi -\varphi ') +\frac{ 2 \pi} {is} 
\eta^{ab} \partial_{\varphi} \delta (\varphi -\varphi ') 
\nonumber\\
&=&\frac{2 \pi} {i s}\left
(D_\varphi\delta(\varphi -\varphi^\prime)\right)^{ab},
\label{poisson}
\end{eqnarray}
which is the $SL(2, {\bf C})$ Kac-Moody algebra in the density form [23, 24]. 
This is an explicit realization of the assumed $SL(2, {\bf C})$ current 
algebra in Ref. [13].
($D_{\varphi}$ is the $\varphi$-th component of the 
covariant derivative $D^{ab}_i=\eta^{ab} \partial_i 
+{\epsilon^{ab}}_c A^c_i$.) [See Ref. [9] for further details.]
Using this Poisson bracket, one finds
\begin{equation}
\{ Q(f), Q(g) \} =Q([f,g]) -\frac{i s}{\pi} \left<A_{\rho} A_{\rho} \right> 
\oint_{\partial D_2} d \varphi ( f^{\rho} \partial_{\varphi} g^{\rho}-
f^{\varphi} \partial_{\varphi} g^{\varphi} ), 
\end{equation}
where $[f,g]^k=f^{\varphi} \partial _{\varphi} g^k -g^{\varphi} \partial _{\varphi} f^k$ is Lie bracket on the boundary circle ($\partial D_2$). 
In general, this algebra does not satisfy the Jacobi identity 
and so the Noether charge $Q(f)$ as a symmetry generator can not be 
accepted. Therefore, the only way to avoid this undesirable situation is to 
consider the subset of transformation with particular 
$f^\rho|_{\partial D_2}\propto \partial _{\varphi} 
{f^{\varphi}|_{\partial D_2}}$
and $g^\rho|_{\partial D_2} \propto \partial _{\varphi} 
g^{\varphi}|_{\partial D_2}$ 
[5, 6, 9] such that only the  third and first order derivatives 
appear in the central term and 
hence (24) satisfies the Jacobi identity [25]: Here, this particular form 
corresponds to the {\it Diff} which deforms $across$ the boundary with 
proportionality to the steepness ($\partial _{\varphi} f^{\varphi}$) of 
{\it Diff} along the circle ($\partial D_2$); The boundary $\partial D_2$ 
responds as an {\it elastic medium} to the deformations. 
Then, (24) will become the Virasoro algebra with central term even at the 
classical level, but with 
an undetermined proportionality constant. This will be determined by 
matching the asymptotic isometries [5 - 7] in the next section.
Before ending this sub-section, I note that the fact of the existness of the 
central term in (24) is a purely Abelian effect which is contained in any 
non-Abelian gauge theories with the non-degenerate 
($\left< t_a t_b \right> \neq 0)$ Lie groups.

\begin{center}
{\bf B. Asymptotic isometries and central charge}
\end{center}

The gauge field (19) has the information about the metric on $\partial D_2$ 
through the relation (15). 
So, the isometries which preserve the metric on $\partial D_2$  can be 
described by {\it Diff} generated by the symmetry generator $Q$. Using (23), 
it is 
found that the transformations of $A_{\varphi}$ who lives on $\partial D_2$, 
generated by $Q(f)$ [9, 26] are
\begin{eqnarray}
\delta_f A _{\varphi}&=& \{ Q(f), A_{\varphi}(\varphi) \} \nonumber \\
&=&D_{\varphi}(f^k A_k), \nonumber \\
\delta_f A_{\rho}&=&0,
\end{eqnarray}
and, using the specific $SL(2, {\bf C})$ gauge field (19), the $\varphi$ part 
of 
(25) becomes
\begin{eqnarray}
\delta_f A_{\varphi} =\frac{1}{2} \left(    
    \begin{array}{cc}
\partial_{\varphi} f^{\rho} & iz~ e^{-\rho} (f^{\rho} -\partial_{\varphi} 
f^{\varphi}) \\
-iz~ e^{\rho} (f^{\rho} +\partial_{\varphi} f^{\varphi}) & 
-\partial_{\varphi} f^{\rho} \end{array}         
\right).
\end{eqnarray}
(25) and (26) are {\it Diff} of gauge fields on $\partial D_2$ regardless of 
the 
radius of $\partial D_2$ and $\rho$ is the proper radius of $\partial D_2$. 
The radius may be finite or infinite; This can be 
even $\rho =0$ which corresponds to the event horizon. Let us consider these 
transformation at infinite boundary 
$\partial D_2$ i.e., $\rho \rightarrow \infty$. Then, in the leading order 
(26) becomes
\begin{eqnarray}
\delta_f A_{\varphi} =\frac{1}{2} \left(    
    \begin{array}{cc}
{\cal O}(1) & {\cal O}(e^{-\rho})  \\
-iz~ e^{\rho} (f^{\rho} +\partial_{\varphi}f_{\varphi}) & 
-{\cal O}(1) \end{array}         
\right),
\end{eqnarray}
where ${\cal O}(1)$ represents the order of $\partial _{\varphi} f^{\rho}$.
Therefore, this transformation (27) gives the isometries $\delta_{f} A_{i}=0$ 
on $\partial D_2$ when
\begin{eqnarray}
f^{\rho}|_{\partial D_2}=-\partial_{\varphi} f^{\varphi}|_{\partial D_2}
\end{eqnarray}
is satisfied. Contrary to the fact of the existness of the central term 
itself, this result is a purely non-Abelian effect which comes from the 
off-diagonal parts. Now, by 
substituting (28) with the insertion of $\left< A_{\rho} A_{\rho} \right>=1/2$
for the black hole solution (19), the algebra (24) becomes the standard 
Virasoro algebra with imaginary number central charge
\begin{eqnarray}
c=-i 24~ s \left < A_{\rho} A_{\rho} \right > =-12 ~i s.
\end{eqnarray}

By defining $Q(f) \equiv \frac{1}{2 \pi} \oint _{\partial D_2} d \varphi 
f^{\varphi}
 (\sum_n L_n e^{-in \varphi})$, $L_n$'s satisfy the momentum space Virasoro 
algebra
\begin{eqnarray}
\{L_m, L_n \} =i (m-n) L_{m+n} +\frac{ic}{12} m (m^2 -1) \delta_{m+n,0}
\end{eqnarray}
with the imaginary value central charge $c$ of (29). In the application of 
this algebra to $Kerr-DS_3$, it is peculiar that the central charge  can be 
completely determined only by considering the exterior region which can not 
be seen 
by an observer moving on a timelike worldline in the interior region. This is 
a physically different situation to $Kerr-AdS_3~
(BTZ)$ [5 - 7].
Moreover, in a general context the central charge might depend on 
$\partial D_2$. But this is impossible because all kinds of $\partial D_2$ 
can be smoothly deformed by {\it Diff} which allows the radial as well as the 
angular deformations, and hence all kinds of $\partial D_2$ are equivalent; 
If the central charge might depend on the radius of circle ($\partial D_2$), 
an absolute length scale must exist in the model but this is contrast to the 
{\it Diff} invariance, which includes the scale invariance of course, of the 
boundary Chern-Simons theory (14)

\begin{center}
{\bf IV. Statistical entropy}
\end{center}

In the computation of the statistical entropy, the zero-mode generators 
$L_0,~\bar{L}_0$ have a crucial role. From the definition, they become
\begin{eqnarray}
L_0 &=&\frac{-i s}{2\pi} \oint _{\partial D_2} \left< A_{\varphi} A_{\varphi}
 +A_{\rho} A_{\rho} \right> = (i l M -J +il ), \nonumber \\
\bar{L}_0&=& (-i l M -J -il ).
\end{eqnarray}
Now, by adjusting the additive constants in $L_0,~\bar{L}_0$ so that they 
vanish
 for the $M=J=0$ case (vacuum solution) [3], one obtains
\begin{eqnarray}
M&=&\frac{1}{il} (L_0 -\bar{L}_0 ), \nonumber \\
J&=&-(L_0 +\bar{L}_0 ).
\end{eqnarray}
Here, there is no condition about the Hermicity for $L_0,~\bar{L}_0$ in 
general to
insure the Hermicity of $M$ and $J$. So, in the general context, one can 
assume that 
$L_0,~\bar{L}_0$ have the complex (eigen)values ${\cal N},~\bar{\cal N}$, 
then $M$ and $J$ become
\begin{eqnarray}
M&=&\frac{1}{i l } ({\cal N}- \bar{\cal N}) =\frac{2}{l} ~{Im} ({\cal N}), 
\nonumber \\
J&=&-({\cal N}+\bar{\cal N}) =-2 ~ {Re} ({\cal N}).
\end{eqnarray}
$M$ and $J$ are controlled by the imaginary and real part of ${\cal N}$, 
respectively.
The eigenvalues ${\cal N}, \bar{\cal N}$ are expressed as
\begin{eqnarray}
{\cal N}=\frac{i}{2}(l M +i J), ~\bar{\cal N} =\frac{-i}{2} (l M -i J).
\end{eqnarray}
There is an argument, by Witten,  of the $unitarity$ 
for the theory with the pure imaginary central charges [18]. But it is not 
well established whether the usual
Cardy's formula [7, 27] for the entropy of a conformal field theory
\begin{eqnarray}
S=2 \pi \sqrt{\frac{c {\cal N}}{6}} +2 \pi \sqrt{\frac{\bar{c} 
\bar{\cal N}}{6}} 
\end{eqnarray}
is valid for the complex valued $c (\bar{c})$ and ${\cal N} (\bar{\cal N})$ 
in general. 
But I will assume this formula and see what happens in my case. Using 
the two main results (29) and (34), one finds the statistical entropy of 
$Kerr-DS_3$ as
\begin{eqnarray}
S=2 \pi \sqrt{l (l M +iJ)} +2 \pi \sqrt{l (l M -iJ)}. \nonumber \\
\end{eqnarray}

For $J=0$ state or a semiclassical regime of large $M,~l$ with small $J$, the
entropy becomes
\begin{eqnarray}
S= 4 \pi l \sqrt{M}.
\end{eqnarray}
This will be the statistical entropy for non-rotating $Kerr-DS_3$ and this 
agrees with the Bekenstein-Hawking's entropy (10) [11, 13]. In this case, 
the state of negative $M$ has no real value entropy which is connected with 
non-existence of the event horizon. From the fact that the result (37) is 
highly sensitive to the central charge $c$ and the desired $c$ is exactly 
obtained by matching isometries at $\rho \rightarrow \infty$, one finds that 
the central charge should be independent on the radius of $\partial D_2$ in 
order to get a consistency with the thermodynamics formula; This fact is 
consistent with {\it Diff} invariance of our theory as I have noted at the 
end of the Sec. III.

On the other hand, when $J\neq 0$, a negative entropy is also a possible 
solution. However, if I assume the smoothness of the entropy change when 
there is 
smooth change from $J=0$, where $S >0$, to $J \neq 0$, it seems to natural to 
consider only the positive entropy solution which agrees exactly with (10). 
In this case, the state of negative $M$ has a real value entropy which is 
connected with the existence of the event horizon. Moreover, for the positive 
entropy  solution, (36) is exactly what can be obtained from BTZ black hole 
entropy (1) by simple replacements: $(i) l \rightarrow i l,~(ii) M \rightarrow
-M $. The part `(i)' corresponds to an analytic continuation and part `(ii)'
is connected to the sign change in $M$ which has been noted below (6). 

From this coincidence, the assumed formula (35) must have some meaning.

\begin{center}
{\bf V. Summary and discussions}
\end{center}

I have considered the $Kerr-DS_3$ space and a statistical evaluation of the 
entropy for the space in the 
$SL(2, {\bf C})$ Chern-Simons gravity formulation. 
It is shown that the space has only one (cosmological) event horizon and 
there is a phase transition between the stable horizon and (unstable) 
evaporating horizon at the 
point $M^2=\frac{1}{3}J^2/l^2$. It is shown also that there is a lower bound 
on the temperature as (13). Then, it is shown that the
Chern-Simons gauge theory with boundaries produces the $SL(2, {\bf C})$ 
Virasoro algebra with imaginary value central term at the $classical$ level; 
In this derivation, it is a basic ingredient that the boundary $\partial D_2$ 
behaves as an {\it elastic medium} to the deformations, i.e., 
$f^\rho|_{\partial D_2}\propto \partial _{\varphi} 
{f^{\varphi}|_{\partial D_2}}$. Using this Virasoro 
algebra, and following the recent approach of BBO, I have 
shown that the statistical entropy for the ${Kerr-DS_3}$ space can be 
calculated by assuming the Cardy's formula (35) even in the imaginary value 
central charge $c$ and complex eigenvalues of generators $L_0,~ \bar{L}_0$. 
This entropy agrees with the Bekenstein-Hawking's formula. My result is 
independent 
on the radius of boundary because of a {\it Diff} invariance of the theory. 
It would be  
interesting to study the Cardy's formula with complex value $c$ and 
${\cal N}$ in a general context 
and understand why it works in my case. It would be also interesting to 
extend to (a) a complex value $c$ which might be related to the inclusion of 
other gauge fields or matter sectors [18], (b) supersymmetric and higher 
dimensional Chern-Simons gravity theories [28], and to understand the analysis of $Kerr-DS_3$ solution within the context of string theory [29].
These remain outstanding challenges.

Note added: After completing this work, I received a paper [30] which 
computes a statistical entropy of the $DS_3$ space (2) using 
$SU(2) \times SU(2)$ Chern-Simons formulation in the Euclidean signature and 
it's result agrees with my result (37) with $M=1$. I thank M. Ortiz for kindly
sending the paper to me before submitting.

\begin{center}
{\bf Acknowledgments}
\end{center}

Most of all, I appreciate Profs. R. Jackiw and P. Oh with whom I had a good 
time to learn about the boundary Chern-Simons theory and symplectic methods 
and other many 
stuffs which became the firm bases of this work. During this work I have 
benefited
from exciting discussions on the black hole physics with J. Cruz and 
stimulating comments by Prof. S. Carlip. I also thank J. Ho, K. Bering, 
Profs. J. Mickelsson, M. Ortiz and Y.-J. Park for helpful correspondence and 
acknowledge the financial support of Korea Research Foundation made in the 
program year 1997. 
This work is supported in part by 
funds provided by the U.S. Department of Energy (D.O.E) under cooperative 
research agreement No. DF-FC02-94ER40818.

\newpage
\begin{center}
{\large \bf References}
\end{center}
\begin{description}

\item{[1]} M. Ba\~nados, Teitelboim and J. Zanelli, Phys. Rev. Lett. {69} 
(1992) 1849; M. Ba\~nados, M. Henneaux, C. Teitelboim and J. Zanelli, Phys. 
Rev. {D48} (1993) 1506.

\item{[2]} S. Carlip, Phys. Rev. {D51} (1995) 632; Phys. Rev. {D55} (1997) 
878.

\item{[3]} A. Strominger, e-print hep-th/9712251.

\item{[4]} J. D. Brown and M. Henneaux, Commun. Math. Phys. {104} (1986) 207.

\item{[5]} M. Ba\~nados, Phys. Rev. {D52} (1996) 5816.

\item{[6]} M. Ba\~nados, T. Brotz, and M. Ortiz, 
e-print hep-th/9802076.

\item{[7]} S. Carlip, e-print hep-th/9806026; M. Ba\`nados and M. Ortiz, 
e-print hep-th/9806089; T. Lee, e-print hep-th/9806113.
 
%

\item{[8]} J. Mickelsson, Lett. Math. Phys. {7} (1983) 45; 
Commun. Math. Phys. {97} (1985) 361.

\item{[9]} P. Oh and M.-I. Park, eprint hep-th/9805178.

\item{[10]} S. Deser and R. Jackiw, Phys. Lett {B139} (1984) 371; 
E. Witten, Commun. Math. Phys. {92} (1984) 455; L. D. Faddeev and 
R. Jackiw, Phys. Rev. Lett. {60} (1988) 1692; L. D. Faddeev, preprint 
HU-TFT-92-5.

\item{[11]} G. W. Gibbons and S. W. Hawking, Phys. Rev. {D15} (1977) 2738.

\item{[12]} S. Deser and R. Jackiw, Ann. Phys. {153} (1984) 405.

\item{[13]} J. Maldacena and A. Strominger, e-print gr-qc/9801096.

\item{[14]} J. D. Brown, J. Creighton and R. B. Mann, Phys. Rev. 
{D 50} (1994) 6394.

\item{[15]} D. Cangemi, M. Leblanc and R. B. Mann, Phys. Rev. {D48} (1993) 3606; D. Bak, D. Cangemi and R. Jackiw, Phys. Rev. {D49} (1994) 5173.

\item{[16]} F. G\"ursey, Ann. Phys. {24} (1963) 211.

\item{[17]} A. Ach\'ucaro and P. K. Townsend, Phys. Lett. {B180} (1986) 89, 
E. Witten, Nucl. Phys. {B311} (1988) 46.

\item{[18]} E. Witten, Commun. Math. Phys. {137} (1991) 29.

\item{[19]} M. Ba\~nados and F. Mendez, eprint hep-th/9806065.

\item{[20]} P. C. W. Davies, Proc. R. Soc. London { A353} (1977) 499; 
Class. Quantum Grav. {6} (1989) 1909.

\item{[21]} D. Pav\'on and J. M. Rub\'i, Phys. Rev. {D37} (1988) 2052; 
R.-G. Cai, Z.-J. Lu and Y.-Z. Zhang, Phys. Rev. {D55} (1997) 853, and 
references therein.

\item{[22]} M.-I. Park, in preparation.

\item{[23]} G. Moore and N. Seiberg, Phys. Lett. {B220} (1989) 422; S. 
Elitzur, G. Moore, A. Schwimer and N. Seiberg, Nucl. Phys. {B326} (1989) 108.

\item{[24]} M. Ba\~nados and A. Gomberoff, Phys. Rev. {D55} (1997) 6162.

\item{[25]} V. G. Kac and A. K. Raina, {Bombay Lecture On  
Highest Weight Representations of Infinite Dimensional Lie Algebras} 
(World Scientific, Singapore, 1987), Sec. 1.3.

\item{[26]} R. Jackiw, Phys. Rev. Lett. {41} (1979) 1635;
Acta. Phys. Austr. Suppl. {XXII} (1980) 383, 
reprinted in  { Diverse Topics in Theoretical and Mathematical Physics}, 
(World Scientific, Singapore, 1995).

\item{[27]} J. A. Cardy, Nucl. Phys. { B270} (1986) 186.

\item{[28]} M. Ba\~nados, K. Boutier, O. Cousaert, M. Henneaux, and M. Ortiz, 
e-print hep-th/9805165; M. Ba\~nados, L. J. Garat and M. Henneaux, Nucl. Phys.
{B476} (1996) 611; T. Lee, e-print/9805182.

\item{[29]} D. Birmingham, I. Sachs and S. Sen, e-print hep-th/9801019.

\item{[30]} M. Ba\~nados, T. Brotz and M. Ortiz, preprint Imperial/TP/97-98, 
DFTUZ 98.
\end{description}
\end{document}